\documentclass[iop]{emulateapj}
 \usepackage{psfrag}
 \usepackage{rotating}
 
 \shorttitle{Mass Loss and Evolution of Polaris.}
 \shortauthors{Neilson et al.}
  
\begin{document}
  
 \title{The period change of the Cepheid Polaris suggests enhanced mass loss}
 \author{Hilding R.~Neilson\altaffilmark{1}}
 \altaffiltext{1}{Argelander Institute for Astronomy, University of Bonn, Auf dem H\"{u}gel 71, 53121 Bonn, Germany}
\email{hneilson@astro.uni-bonn.de}
\author{Scott G.~Engle\altaffilmark{2}}
\altaffiltext{2}{Department of Astronomy \& Astrophysics, Villanova University, 800 Lancaster Ave. Villanova, PA 19085 USA}
 \author{Ed Guinan\altaffilmark{2}}
\author{Norbert Langer\altaffilmark{1}}
\author{Richard P. Wasatonic\altaffilmark{2}}
\author{David B. Williams\altaffilmark{3}}
\altaffiltext{3}{American Association of Variable Star Observers, 49 Bay State Road, Cambridge, MA 02138 USA}

\begin{abstract}
Polaris is one of the most observed stars in the night sky, with recorded observations spanning more than 200 years.  From these observations, one can study the real-time evolution of Polaris via the secular rate of change of the pulsation period.  However, the measurements of the rate of period change do not agree with predictions from state-of-the-art stellar evolution models.  We show that this may imply that Polaris is currently losing mass at a rate of $\dot{M} \approx 10^{-6}~M_\odot$~yr$^{-1}$ based on the difference between modeled and observed rates of period change, consistent with pulsation-enhanced Cepheid mass loss. A relation between the rate of period change and mass loss has important implications for understanding stellar evolution and pulsation, and provides insight into the current Cepheid mass discrepancy.
 
\end{abstract}
\keywords{stars: individual (Polaris) --- stars: evolution --- stars: variables: Cepheids ---  stars: mass-loss}

\section{Introduction}
Classical Cepheids are radially pulsating variable stars and are ideal laboratories for understanding stellar astrophysics and evolution. Radial pulsation constrains stellar structure along with Cepheid masses and luminosities via the Period-Mean Density relation.   However, observations of the rate of pulsation period change directly constrain stellar evolution because this measures the change of the stellar mean density as a function of time.  

\cite{Turner2006} compiled measurements of rates of period change for a sample of Galactic Cepheids and showed that they broadly agree with rates of period change as a function of pulsation period computed from stellar evolution models. The authors also found that observed rates of period change for Polaris and DX Gem are larger than those of most Cepheids and are consistent with being a Cepheid on its first crossing of the instability strip.  The first crossing occurs as a star evolves across the Hertzsprung gap after main sequence evolution on a Kelvin-Helmholtz timescale.  

That work illustrated the power of measured rates of period change to constrain stellar evolution models, however, one can go a step further.  By combining observations of the rate of period change with measurements of mean stellar radius, mean luminosity and/or effective temperature, we can directly test the input physics used in stellar evolution models.  In this work, we produce a grid of state-of-the-art stellar evolution models and compare stellar evolution tracks with the measured radius and effective temperature of the first-overtone Cepheid Polaris.  For evolutionary tracks that cross the Cepheid instability strip, we compute theoretical rates of period change for comparison.  From this comparison, we constrain the physics of stellar evolution models and other fundamental properties of Polaris.

Polaris is an ideal candidate for this analysis. It is an astrometric binary (as part of a multiple star system) with a dynamically measured mass of $M=4.5^{+2.2}_{-1.4}~M_\odot$ \citep{Kamper1996, Evans2008}, has been observed using K-band interferometry yielding a measurement of the its mean angular diameter \citep{Merand2006}, and has a precise, revised Hipparcos parallax measurement \citep{Leewen2007}.  The effective temperature of Polaris has also been measured spectroscopically \citep{Usenko2005}.  The combination of these observations provide strong constraints for stellar evolution models.   There have also been a number of measurements of the rate of period change for Polaris based on the combination of recent and archival observations. \cite{Spreckley2008} measured a rate of period change $\dot{P} = 4.90\pm 0.26~$s~yr$^{-1}$ based on 50 years of observations, while \cite{Turner2005} found a rate of $\approx4.4~$s~yr$^{-1}$ based on 160 years of observations.  In this work, we present a rate of period change based on data from \cite{Turner2005} with new data added from measurements over the past decade.

In the next section, we describe the measurements of the rate of period change in greater detail. In Section 3, we describe the stellar evolution code and models used in this study, while in Section 4, we compare the predicted rates of period change with the observed rate where we find disagreement.  In Section 5, we discuss the implications of our study toward understanding the structure and evolution of Cepheids as well as the problem of the Cepheid mass discrepancy.

\section{Observed Rate of Period Change}
\begin{figure}[t]
\begin{center}
\plotone{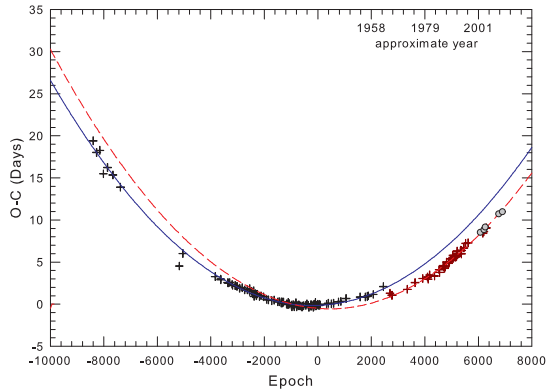}
\end{center}
\caption{O-C data from 1844/5-2011, from both radial velocity and photometry measures, are shown for the Cepheid Polaris. For the purposes of fitting, the data were divided into two sets (pre-1963 and post-1965) because of the apparent discontinuity in the O-C curve around 1963-66 (Epoch 2400-2600). The weighted, least squares quadratic fits to the data sets are also shown, where the blue solid line is the fit to the earlier dataset and the red dashed line is the fit to the later. The crosses represent data from \cite{Turner2005} and circles our new timing data.  The early data show a period increase of $4.44\pm0.19$~s~yr$^{-1}$ and the later data show a period increase of $4.50\pm1.45 $~s~yr$^{-1}$.}\label{fig0}
\end{figure}

The recent comprehensive study by  \cite{Turner2005} of photometric and radial velocity measurements (up to 2004) of Polaris yielded rates of increase in the pulsation period of Polaris of $\dot{P} = 4.45\pm0.03$ and $4.28\pm0.73$~s~yr$^{-1}$. This result is in good accord with previous period studies of Polaris. We added five photometrically determined timings (covering 2002-2011) to those given in Table 1 of  \cite{Turner2005}.  The resulting O-C analysis plot is given in Figure \ref{fig0}.  As noted previously, there is an apparent discontinuity in the O-C plot occurring near epoch 2400-2600. Analyses of the timings prior to (1844/5- November, 1962) and after (July, 1965-2011) this small anomaly yielded common values of period increase Ð $\dot{P} = 4.44\pm0.19$ and $4.50\pm1.45$~s~yr$^{-1}$, respectively (the larger uncertainty for the later period change is due to the more limited time span covered). When all data are taken as one set, the calculated period increase is $3.93\pm0.08$~s~yr$^{-1}$. For the purpose of this study, we adopt the value of $\dot{P}= 4.47\pm1.46$~s~yr$^{-1}$, as the average of the two fits on the two separate sets described above. In fact, as noted by  \cite{Turner2005}, the discontinuity in the O-C curve coincides with the possible beginning of an accelerated decrease in the light amplitude of Polaris, however this requires further study.

\section{Stellar Evolution Models}
We compute state-of-the-art stellar evolution models using the \cite{Yoon2005} version of the code developed by \cite{Heger2000}. Models are computed for the mass range of $3$ - $6.8~M_{\odot}$ in steps of $0.1~M_{\odot}$, consistent with the measured mass of Polaris determined by \cite{Evans2008}. We also vary the amount of convective core overshooting.  Convective core overshooting is parameterized as the distance that convective eddies penetrate above the core boundary in a star over its evolutionary timescale.  The distance is defined, in the \cite{Yoon2005} code, to be $\alpha_{c} H_{P}$, where $H_{P}$ is the pressure scale height and $\alpha_{c}$ is a free parameter.  In this work, we compute models for $\alpha_{c}=0$-$0.4$ in steps of $0.1$.  We vary the amount of overshooting because overshooting during main sequence stellar evolution acts to produce a more massive helium core and hence a more luminous Cepheid for the same stellar mass with no overshooting.  There is no enhanced Cepheid mass loss assumed like that explored by \cite{Neilson2011} and the models are assumed to have the \cite{Grevesse1998} solar composition.  Mass loss is treated using the \cite{Jager1988} recipe for cool stars and the \cite{Kudritzki1989} prescription for hot stars.

\begin{figure*}[t]
\begin{center}
\plottwo{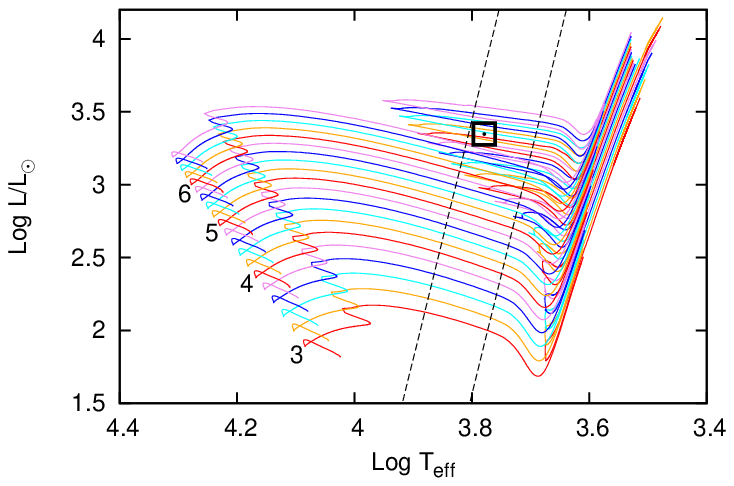}{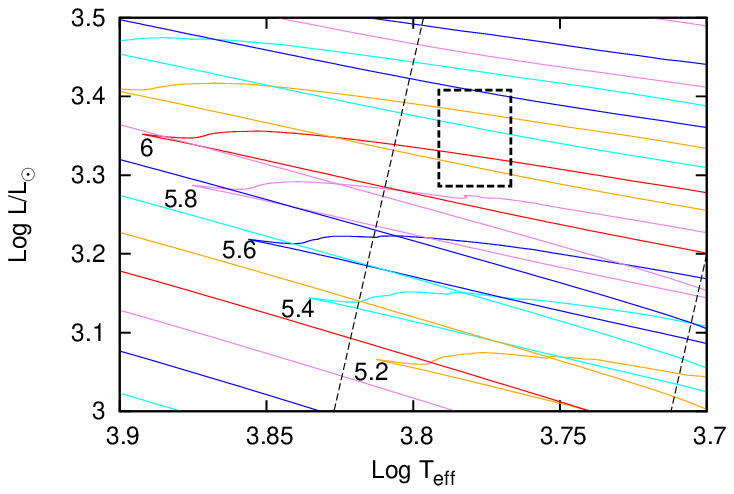}
\end{center}
\caption{Stellar evolution tracks computed using the \cite{Yoon2005} code for initial masses ranging from $3~M_\odot$ to $6.8~M_\odot$, assuming no convective core overshooting, $\alpha_c = 0$ (left) and the region of the HR diagram containing the effective temperature and luminosity of Polaris(right). The black dashed lines represent the boundaries of the Cepheid instability strip \citep{Bono2000} while the boxes denote the effective temperature and luminosity of Polaris.}\label{fig1}
\end{figure*}
Evolutionary tracks for models without convective core overshoot are shown in Fig.~\ref{fig1} for a sample of stellar masses along with the measured effective temperature and luminosity of Polaris.  The effective temperature was spectroscopically determined by \cite{Usenko2005} to be $T_{eff}=6015\pm170~K$, while the radius is $R=43.5\pm 0.8~R_\odot$ based on angular diameter measurements \citep{Merand2006} and the revised Hipparcos parallax \citep{Leewen2007}, hence $\log L/L_\odot = 3.347\pm 0.061$.  In Fig.~\ref{fig1}, not all models are consistent with the observed properties, suggesting the observations are precise enough to constrain models.  In the next section, we compute the rates of period change from the stellar evolution models.
  
\section{Rates of Period Change}
\cite{Turner2006} derived a relation for computing the rate of period change from the predicted stellar properties of evolution models from the Period-Mean Density relation and noting that the pulsation constant $Q \propto P^{1/8}$ \citep{Fernie1967},
\begin{equation}\label{e1}
\frac{\dot{P}}{P} = \frac{6}{7}\frac{\dot{L}}{L} - \frac{24}{7}\frac{\dot{T}_{\rm{eff}}}{T_{\rm{eff}}}.
\end{equation}
This relation assumes that mass loss is insignificant.  We compute rates of period change $\dot{P}$ using this relation where the values of $L$ and $T_{\rm{eff}}$ along with the time derivatives are given by the stellar evolution models.  

\begin{figure}[t]
\begin{center}
\plotone{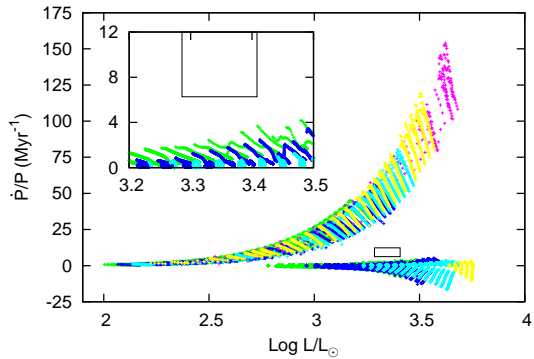}
\end{center}
\caption{Predicted rates of period change scaled by pulsation period as a function of stellar luminosity computed from the stellar evolution models when the evolution tracks intersect assumed boundaries of the Cepheid instability strip \citep[see][for details]{Bono2000}.  The different colors represent differing amount of assumed convective core overshooting, $\alpha_c = 0$ (green), $0.1$ (blue), $0.2$ (light blue), $0.3$ (yellow), and $0.4$ (magenta). The box represents the observed luminosity and the rate of period change range for Polaris. The inset zooms in on the luminosity range for Polaris and shows a distinct difference between theoretical and observed rates of period change.}\label{f1}
\end{figure}
\begin{figure}[t]

\begin{center}
\plotone{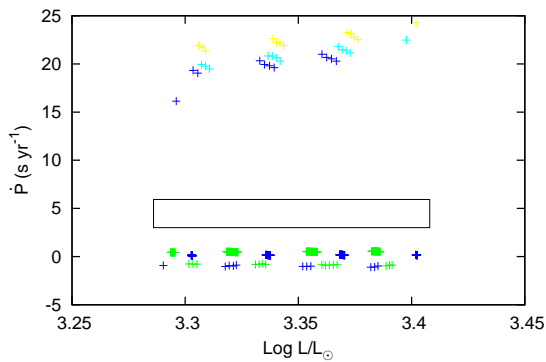}
\end{center}
\caption{Predicted rates of period change as a function of stellar luminosity computed from stellar evolution models when the  models have effective temperature and radius consistent with observations. The colors have the same meaning as in Fig.~\ref{f1}.}\label{f2}
\end{figure}

When the stellar evolution models predict radii and effective temperatures that fall within the boundaries of the Cepheid instability strip, we compute the model relative rate of period change, ${\dot{P}}/{P}$ using Eq.~1.  This is done to avoid making any assumptions of the pulsation period from the theoretical stellar models. We show the predicted rates of period change as a function of stellar luminosity in Fig.~\ref{f1}. The observed relative rate of period change is scaled by the calculated fundamental period for Polaris, $P_0 = 5.61$~days.  The calculated fundamental period is the overtone period plus $0.15~$dex \citep{Turner2005}.   

We find that evolutionary models with masses $5.4$ - $6.4~M_\odot$ and convective core overshooting $\alpha_c \le 0.4$ predict effective temperatures and radii consistent with observations, but the predicted mass is degenerate with the assumed amount of overshooting.  This can be seen in Fig.~\ref{fig1} for models with $\alpha_c = 0$, and masses, $M = 5.9$ - $6.4~M_\odot$; smaller masses are measured for $\alpha_c>0$. The degeneracy between mass and $\alpha_c$ is discussed in detail by \cite{Keller2008}. Increasing  $\alpha_c$ in stellar evolution models decreases the blue loop width, however, this does not change the rate of period change for a given stellar luminosity during blue loop evolution. The rate of period change is dominated by the most important timescale at each stage of stellar evolution. On the Hertzsprung gap, a star undergoes thermal relaxation, meaning the timescale and the rate of period change is defined by the Kelvin timescale $\tau_{\rm{K}}$, which is $\propto M^2$.  The evolutionary timescale of a star on the blue loop is defined by the helium nuclear burning timescale, $\tau_{\rm{He}}$, which itself depends on the helium core mass $M_{\rm{He}}$, i.e. the luminosity.  The helium core mass is determined by the combination of stellar mass and the amount of assumed convective core overshoot.  Thus, the rate of period change for an evolution model with a given luminosity, mass, and no overshooting is approximately the same as for a model with the same luminosity, but a different amount of overshooting and corresponding mass.

The rates of period change are plotted in Fig.~\ref{f2} as a function of luminosity for the models that have a predicted radius and effective temperature consistent with the observed values during the blue loop evolution. This plot differs from the results in Fig.~\ref{f1} such that the theoretical value of the pulsation period is assumed to be that of Polaris.  Any uncertainty introduced by this assumption will not affect the results.   The stellar evolution models predict rates of period change that occur in three regions for the plot in Fig.~\ref{f1}.  The first region includes rates of period change where $\dot{P} > 15$~s~yr$^{-1}$ in Fig~\ref{f2}, which corresponds to stellar models rapidly evolving across the Hertzsprung gap.  The second region is for $\dot{P} < 0$~s~yr$^{-1}$, that corresponds to blueward evolution during the blue loop phase of stellar evolution, while the third region is $0<\dot{P} \le 1$~s~yr$^{-1}$ corresponding to redward evolution during the blue loop phase.   

The observed rate of period change is inconsistent with any of the three phases of stellar evolution, and contrary to previous claims Polaris cannot be a first-crossing Cepheid \citep{Turner2005, Turner2006}. This difference was also found by \cite{Spreckley2008}. The question is what physics is missing that can resolve this difference. We rewrite the rate of period change from Eq.~\ref{e1} to be
\begin{equation}\label{pmdot}
\frac{\dot{P}}{P} = -\frac{4}{7}\frac{\dot{M}}{M} + \frac{6}{7}\frac{\dot{L}}{L} - \frac{24}{7}\frac{\dot{T}_{\rm{eff}}}{T_{\rm{eff}}},
\end{equation}
where we assume that the change of mass provides a non-negligible contribution to the rate of period change \citep{Turner2006, Neilson2008}. Since Polaris has not been observed accreting material then the rate of change of stellar mass is due to mass loss only and is always $<0$. This means that mass loss only acts to increase the rate of period change.  

If we consider the largest value for the rate of period change during the blue loop evolution in Fig.~\ref{f2}, $\dot{P}  \approx 1~$s~yr$^{-1}$, then to resolve the difference between the theoretical and observed rates of period change the rate of change of stellar mass must be $\le -7\times 10^{-6}~M_\odot$~yr$^{-1}$.  We note this result ignores the possibility for feedback on the stellar radius due to mass loss, but the feedback is expected to decrease the mass-loss rate by less than an order of magnitude. 

\section{Discussion}
We find that the observed rate of period change, from all sources, is inconsistent with predictions from stellar evolution models unless one includes significant mass loss. Previous evidence for Cepheid mass loss is based on observations of infrared excess \citep{Neilson2009, Neilson2010, Marengo2010, Barmby2011}, as well as  21-cm radio observations of a nebula surrounding $\delta$ Cephei \citep{Matthews2011}.
These previous observations did not confirm mass loss because any circumstellar material could be relics from earlier stages of stellar evolution or material coincidently existing along the observer's line-of-sight.  Thus, the rate of period change provides a unique and strong argument for enhanced mass loss in Cepheids.

The measured mass-loss rate for Polaris is similar to that found from the 21cm radio ro observations of $\delta$ Cephei, as well as measurements from infrared excesses, depending on the assumed dust properties.  \cite{Neilson2008,Neilson2009b} predicted theoretical mass-loss rates for Galactic Cepheids that are significantly less  than $10^{-6}~M_\odot~$yr$^{-1}$, but stellar evolution models including the same pulsation-driven mass loss theory \citep{Neilson2011} found rates of the order $10^{-7}$ to $10^{-6}~M_\odot~$yr$^{-1}$. That work suggests that Cepheid mass loss may solve the Cepheid mass discrepancy assuming some moderate convective core overshooting.   The significant amount of mass loss for Polaris implies that Cepheid pulsation is a wind driving mechanism.  

The measured rate of period change for Polaris is just one piece of evidence contributing to the understanding of Cepheid mass loss.  Recent X-ray \citep{Evans2010} and UV \citep{Engle2009} observations of Polaris indicate the presence of hot $T \approx 10^5~$K plasma in the envelope, which may originate from shocks generated by pulsation in the Cepheid \citep{Fokin1996}.  Further spectral observations at near-UV wavelengths could provide additional constraints on the mass loss mechanism in Cepheids.

Polaris may not be unique as a Cepheid with significant mass loss. Large rates of period change are considered to be a property of first overtone pulsating Cepheids \citep{Szabados1983, Evans2002},  implying that many of these Cepheids, if not all, have significant mass-loss rates during the blue loop stage of stellar evolution.   \cite{Turner2010} found that four Cepheids are first overtone or double mode pulsators and argued they must be evolving along the Hertzsprung gap, based on the rates of period change.  These are Polaris, DX Gem, BY Cas, and HDE 344787. However, we find that stellar evolution fits to the observed radius and effective temperature of Polaris suggest a rate of period change for the first crossing that is about four times larger than the observed rate; hence Polaris is most likely evolving along the blue loop.  The observed rate of period change for Polaris is similar to that of BY Cas and DX Gem, while for HDE 344787, $\dot{P} = 12.96\pm 2.41~$s~yr$^{-1}$, suggesting that the rates for BY Cas and DX Gem may also be consistent with extreme mass loss for evolution along the blue loop.  Observations constraining the radius, temperature, and/or luminosity along with rates of period change will test if all first overtone Cepheids have extreme mass loss. 

The results  also suggest that more time-series observations of Polaris in multiple wavelength regimes will help constrain the mass-loss history and mass-loss mechanism for this star and unravel a mystery of the North Star.

\acknowledgements
HRN is grateful for funding from the Alexander von Humboldt Foundation, and to Dr. Matteo Cantiello for helpful conversations regarding stellar evolution calculations. EG and SE are grateful for the support from NASA Grants HST-GO-11726.01, HST-GO-12302.01, NNX08AX37G and NASA/JPL Grant No. 40968.

 \bibliographystyle{apj}

\begin{thebibliography}{29}
\expandafter\ifx\csname natexlab\endcsname\relax\def\natexlab#1{#1}\fi

\bibitem[{{Barmby} {et~al.}(2011){Barmby}, {Marengo}, {Evans}, {Bono},
  {Huelsman}, {Su}, {Welch}, \& {Fazio}}]{Barmby2011}
{Barmby}, P., {Marengo}, M., {Evans}, N.~R., {Bono}, G., {Huelsman}, D., {Su},
  K.~Y.~L., {Welch}, D.~L., \& {Fazio}, G.~G. 2011, \aj, 141, 42

\bibitem[{{Bono} {et~al.}(2000){Bono}, {Castellani}, \& {Marconi}}]{Bono2000}
{Bono}, G., {Castellani}, V., \& {Marconi}, M. 2000, \apj, 529, 293

\bibitem[{{de Jager} {et~al.}(1988){de Jager}, {Nieuwenhuijzen}, \& {van der
  Hucht}}]{Jager1988}
{de Jager}, C., {Nieuwenhuijzen}, H., \& {van der Hucht}, K.~A. 1988, \aaps,
  72, 259

\bibitem[{{Engle} {et~al.}(2009){Engle}, {Guinan}, {Depasquale}, \&
  {Evans}}]{Engle2009}
{Engle}, S.~G., {Guinan}, E.~F., {Depasquale}, J., \& {Evans}, N. 2009, in
  American Institute of Physics Conference Series, Vol. 1135, American
  Institute of Physics Conference Series, ed. {M.~E.~van Steenberg,
  G.~Sonneborn, H.~W.~Moos, \& W.~P.~Blair }, 192--197

\bibitem[{{Evans} {et~al.}(2010){Evans}, {Guinan}, {Engle}, {Wolk}, {Schlegel},
  {Mason}, {Karovska}, \& {Spitzbart}}]{Evans2010}
{Evans}, N.~R., {Guinan}, E., {Engle}, S., {Wolk}, S.~J., {Schlegel}, E.,
  {Mason}, B.~D., {Karovska}, M., \& {Spitzbart}, B. 2010, \aj, 139, 1968

\bibitem[{{Evans} {et~al.}(2002){Evans}, {Sasselov}, \& {Short}}]{Evans2002}
{Evans}, N.~R., {Sasselov}, D.~D., \& {Short}, C.~I. 2002, \apj, 567, 1121

\bibitem[{{Evans} {et~al.}(2008){Evans}, {Schaefer}, {Bond}, {Bono},
  {Karovska}, {Nelan}, {Sasselov}, \& {Mason}}]{Evans2008}
{Evans}, N.~R., {Schaefer}, G.~H., {Bond}, H.~E., {Bono}, G., {Karovska}, M.,
  {Nelan}, E., {Sasselov}, D., \& {Mason}, B.~D. 2008, \aj, 136, 1137

\bibitem[{{Fernie}(1967)}]{Fernie1967}
{Fernie}, J.~D. 1967, \aj, 72, 1327

\bibitem[{{Fokin} {et~al.}(1996){Fokin}, {Gillet}, \&
  {Breitfellner}}]{Fokin1996}
{Fokin}, A.~B., {Gillet}, D., \& {Breitfellner}, M.~G. 1996, \aap, 307, 503

\bibitem[{{Grevesse} \& {Sauval}(1998)}]{Grevesse1998}
{Grevesse}, N. \& {Sauval}, A.~J. 1998, \ssr, 85, 161

\bibitem[{{Heger} {et~al.}(2000){Heger}, {Langer}, \& {Woosley}}]{Heger2000}
{Heger}, A., {Langer}, N., \& {Woosley}, S.~E. 2000, \apj, 528, 368

\bibitem[{{Kamper}(1996)}]{Kamper1996}
{Kamper}, K.~W. 1996, \jrasc, 90, 140

\bibitem[{{Keller}(2008)}]{Keller2008}
{Keller}, S.~C. 2008, \apj, 677, 483

\bibitem[{{Kudritzki} {et~al.}(1989){Kudritzki}, {Pauldrach}, {Puls}, \&
  {Abbott}}]{Kudritzki1989}
{Kudritzki}, R.~P., {Pauldrach}, A., {Puls}, J., \& {Abbott}, D.~C. 1989, \aap,
  219, 205

\bibitem[{{Marengo} {et~al.}(2010){Marengo}, {Evans}, {Barmby}, {Matthews},
  {Bono}, {Welch}, {Romaniello}, {Huelsman}, {Su}, \& {Fazio}}]{Marengo2010}
{Marengo}, M., {Evans}, N.~R., {Barmby}, P., {Matthews}, L.~D., {Bono}, G.,
  {Welch}, D.~L., {Romaniello}, M., {Huelsman}, D., {Su}, K.~Y.~L., \& {Fazio},
  G.~G. 2010, \apj, 725, 2392

\bibitem[{{Matthews} {et~al.}(2011){Matthews}, {Marengo},
  {Evans}, \& {Bono}}]{Matthews2011}
{Matthews}, L.~D., {Marengo}, M., {Evans}, N.~R., \& {Bono}, G.  2011, ApJ, in press, 
arXiv:1112.0028

\bibitem[{{M{\'e}rand} {et~al.}(2006){M{\'e}rand}, {Kervella}, {Coud{\'e} du
  Foresto}, {Perrin}, {Ridgway}, {Aufdenberg}, {ten Brummelaar}, {McAlister},
  {Sturmann}, {Sturmann}, {Turner}, \& {Berger}}]{Merand2006}
{M{\'e}rand}, A., {Kervella}, P., {Coud{\'e} du Foresto}, V., {Perrin}, G.,
  {Ridgway}, S.~T., {Aufdenberg}, J.~P., {ten Brummelaar}, T.~A., {McAlister},
  H.~A., {Sturmann}, L., {Sturmann}, J., {Turner}, N.~H., \& {Berger}, D.~H.
  2006, \aap, 453, 155

\bibitem[{{Neilson} {et~al.}(2011){Neilson}, {Cantiello}, \&
  {Langer}}]{Neilson2011}
{Neilson}, H.~R., {Cantiello}, M., \& {Langer}, N. 2011, \aap, 529, L9+

\bibitem[{{Neilson} \& {Lester}(2008)}]{Neilson2008}
{Neilson}, H.~R. \& {Lester}, J.~B. 2008, \apj, 684, 569

\bibitem[{{Neilson} \& {Lester}(2009)}]{Neilson2009b}
---. 2009, \apj, 690, 1829

\bibitem[{{Neilson} {et~al.}(2009){Neilson}, {Ngeow}, {Kanbur}, \&
  {Lester}}]{Neilson2009}
{Neilson}, H.~R., {Ngeow}, C.-C., {Kanbur}, S.~M., \& {Lester}, J.~B. 2009,
  \apj, 692, 81

\bibitem[{{Neilson} {et~al.}(2010){Neilson}, {Ngeow}, {Kanbur}, \&
  {Lester}}]{Neilson2010}
---. 2010, \apj, 716, 1136

\bibitem[{{Spreckley} \& {Stevens}(2008)}]{Spreckley2008}
{Spreckley}, S.~A. \& {Stevens}, I.~R. 2008, \mnras, 388, 1239

\bibitem[{{Szabados}(1983)}]{Szabados1983}
{Szabados}, L. 1983, \apss, 96, 185

\bibitem[{{Turner} {et~al.}(2006){Turner}, {Abdel-Sabour Abdel-Latif}, \&
  {Berdnikov}}]{Turner2006}
{Turner}, D.~G., {Abdel-Sabour Abdel-Latif}, M., \& {Berdnikov}, L.~N. 2006,
  \pasp, 118, 410

\bibitem[{{Turner} {et~al.}(2010){Turner}, {Majaess}, {Lane}, {Percy},
  {English}, \& {Huziak}}]{Turner2010}
{Turner}, D.~G., {Majaess}, D.~J., {Lane}, D.~J., {Percy}, J.~R., {English},
  D.~A., \& {Huziak}, R. 2010, Odessa Astronomical Publications, 23, 125

\bibitem[{{Turner} {et~al.}(2005){Turner}, {Savoy}, {Derrah}, {Abdel-Sabour
  Abdel-Latif}, \& {Berdnikov}}]{Turner2005}
{Turner}, D.~G., {Savoy}, J., {Derrah}, J., {Abdel-Sabour Abdel-Latif}, M., \&
  {Berdnikov}, L.~N. 2005, \pasp, 117, 207

\bibitem[{{Usenko} {et~al.}(2005){Usenko}, {Miroshnichenko}, {Klochkova}, \&
  {Yushkin}}]{Usenko2005}
{Usenko}, I.~A., {Miroshnichenko}, A.~S., {Klochkova}, V.~G., \& {Yushkin},
  M.~V. 2005, \mnras, 362, 1219

\bibitem[{{van Leeuwen} {et~al.}(2007){van Leeuwen}, {Feast}, {Whitelock}, \&
  {Laney}}]{Leewen2007}
{van Leeuwen}, F., {Feast}, M.~W., {Whitelock}, P.~A., \& {Laney}, C.~D. 2007,
  \mnras, 379, 723

\bibitem[{{Yoon} \& {Langer}(2005)}]{Yoon2005}
{Yoon}, S.-C. \& {Langer}, N. 2005, \aap, 443, 643

\end{thebibliography}

\end{document}